\documentclass[preprint,review,12pt]{elsarticle}
\usepackage{graphicx}
\usepackage{epsf}
\usepackage{amsmath}
\usepackage{lscape}
\usepackage{color}
\usepackage{dcolumn}
\usepackage{amsmath}
\usepackage{rotating}
\usepackage{soul}
\usepackage{longtable}

\newcommand{\cm}{cm$^{-1}$}

\newcommand{\ai}{\textit{ab initio }}

\journal{Journal of Quantitative Spectroscopy \& Radiative Transfer}

\begin{document}

\begin{frontmatter}

\title{An experimental water line list at 1950 K in the 6250 -- 6670 \cm\ region  }
\author{Lucile Rutkowski, Aleksandra Foltynowicz}
\address{Department of Physics, Ume\aa\ University, 901 87 Ume\aa, Sweden}
\author{Florian M. Schmidt}
\address{Thermochemical Energy Conversion Laboratory, Department of Applied Physics and Electronics, Ume\aa\ University, 901 87 Ume\aa, Sweden}
\author{Alexandra C. Johansson, Amir Khodabakhsh}
\address{Department of Physics, Ume\aa\ University, 901 87 Ume\aa, Sweden}
\author{ Aleksandra A. Kyuberis, Nikolai F. Zobov}
\address{Institute of Applied Physics, Russian Academy of Sciences,
Ulyanov Street 46, Nizhny Novgorod, Russia 603950.}
\author{ Oleg L. Polyansky, Sergei N. Yurchenko, Jonathan Tennyson}
\address{Department of Physics and Astronomy, University College London,
London, WC1E 6BT, UK}

\begin{abstract}
  An absorption spectrum of H$_2$$^{16}$O at 1950 K is recorded in a premixed
  methane/air flat flame using a cavity-enhanced optical frequency comb-based Fourier
  transform spectrometer. 2417 absorption lines are identified in the
  6250 -- 6670 \cm\ region with an accuracy of about 0.01 \cm.
  Absolute line intensities are retrieved using temperature and
  concentration values obtained by tunable diode laser
  absorption spectroscopy.  Line assignments are made using a
  combination of empirically known energy levels and predictions from
  the new POKAZATEL variational line list. 2030 of the observed lines
  are assigned to 2937 transitions, once blends are taken into
  account. 126 new energy levels of H$_2$$^{16}$O are identified.
 The assigned transitions belong to 136 bands and span rotational
  states up to $J=27$.
%\red{We should also mention MARVEL and comparison to HITRAN in the abstract. Should we write in past tense instead of present? That would sound better for the experimental part.}

% MARVEL data is mentioned and there is no need to mention HITRAN
\end{abstract}

\begin{keyword}

water \sep absorption \sep Fourier transform spectroscopy,
optical cavity
\sep frequency comb \sep \ai calculations

\end{keyword}
\end{frontmatter}

\section{Introduction}
Water is ubiquitous and its spectrum is important for a whole range of
terrestrial and astronomical applications. Serious attempts have been made
to characterize the spectrum of hot water both
experimentally by observation of spectra
\cite{83PiCoCaFl,91PeAnHeDe,jt203,jt205,jt211,jt212,98EsWaHoRo,jt231,jt251,jt297,04CoPiVeLa,jt348,jt362,jt377,jt437,12YuPeDrMa,14CoMaPi}
and theoretically by the computation of extensive line lists
\cite{jt143,92WaRoxx.CO2,jt197,ps97,JJS01,jt378,jtpoz}. These line
lists are used to inform databases concerned with models of hot bodies
such as HITEMP \cite{jt480} and ExoMol \cite{jt528,jt631}. A
comprehensive assessment of water spectroscopy was undertaken by an
IUPAC task group \cite{jt539,jt562} whose work is currently being
updated \cite{jtwaterupdate}.

The ubiquity of water means that understanding its spectrum at all
wavelengths and temperatures is always important. The
spectrum of hot water is of particular interest in regions were
absorption by room temperature water is weak. The present work
concentrates on one such region as it probes the spectrum of hot water
in the conventional telecom window (1.53 -- 1.565 $\mu$m) as well as
the astronomers H-band (1.5 -- 1.8 $\mu$m). These regions are
useful for remote sensing of hot water spectra due to
the reduced atmospheric absorption. Previous high-temperature water spectra analyzed for this region
\cite{76FlCaMa,77CaFlMa,92MaDaCaFl,92DaMaCaFla,02MiTyStAl,jt297,jt348,jt362}
were recorded in emission in flames at atmospheric pressure at moderate spectral
resolution; in addition, due to the lack of thermal stability, these
spectra did not provide usable information on the line intensities.

This paper presents a high temperature water absorption spectrum
measured at Ume\aa\ University. The spectrum is measured in a
premixed methane/air flat flame at atmospheric pressure using a
cavity-enhanced optical frequency comb-based Fourier transform
spectrometer (FTS) \cite{UMU3}. The combination of an FTS with a
frequency comb allows the measurement of broadband and high resolution
molecular spectra in short acquisition times and without visible influence
of the instrumental line shape \cite{UMU9,UMU1}, while the cavity
provides high sensitivity to absorption \cite{UMU5}. The ability to
measure the present spectrum simultaneously over a broad bandwidth
reduces systematic errors and the influence of fluctuations of the
environmental conditions. The spectrum is recorded at high resolution
(0.033 \cm) in the near-infrared 6250 -- 6670 \cm\ region, and line
positions are identified with an accuracy of 0.01 \cm. Knowledge of
the temperature and water concentration, which have previously been
measured for that specific burner by Qu {\it et al.} \cite{UMU2} using
tunable diode laser absorption spectroscopy, as well as the thermally
stable conditions, allow absolute line intensities to be determined.

The measured absorption spectrum is compared to the newly computed
POKAZATEL hot line list \cite{jtpoz} augmented by the inclusion of
empirical energy levels \cite{jt539}.  This comparison
allows us to assess both the
contents of the measured spectrum and the reliability of the computed
line list. The POKAZATEL
line list is then used to make assignments to the spectrum resulting
in a significant number of newly identified transitions and energy
levels.  

The following two sections of the paper describe the
experimental set-up and results. Section 4 presents the experimental
water line list.  Comparisons with the computed line lists,
particularly the most recent one \cite{jtpoz}, are given in section 5,
followed by conclusions in section 6.
% \red{Mention comparison to HITRAN?}

\section{Experimental setup}

The experimental setup is described in detail in references
\cite{UMU3,UMU10} and is therefore only briefly summarized here. The
spectrometer consists of an Er:fiber femtosecond laser with a
repetition rate of 250 MHz (0.0083 cm$^{-1}$), a 60 cm long enhancement cavity
with a
finesse of around 150, and a fast-scanning Fourier transform
spectrometer (FTS). The comb is locked to the cavity using the
two-point Pound-Drever-Hall method \cite{UMU5,UMU6} with locking
points at 6330 and 6450 cm$^{-1}$.
The cavity is open to air, and a flat flame
burner \cite{UMU7} is placed in its center. The burner is operated on premixed
methane/air at stoichiometric ratio with a total flow rate of 10 L/min. The comb
beam probes the line of sight in the flame (flame diameter of 3.8 cm) at
atmospheric pressure and at a height above the burner (HAB) of 2.5 mm. At this HAB
the temperature and species are rather homogeneously distributed along the line
of sight \cite{UMU2,UMU7}, the average flame temperature is $1950\pm 50$ K, the
average water concentration is $17\pm 1$\% (both characterized using tunable
diode laser absorption spectroscopy  \cite{UMU2}), and the average hydroxide
(OH) concentration is 0.28\% \cite{UMU10}.

The light transmitted through the cavity is coupled into an optical fiber
connected to the input of a fast-scanning FTS with an auto-balancing detector
that acquires a spectrum with 0.033 cm$^{-1}$ resolution in 0.4 s. The optical
path difference is calibrated using a stabilized HeNe laser whose beam is
co-propagating with the comb beam in the FTS. The wavelength of the HeNe laser
is calibrated by comparing the positions of the OH lines in the spectrum to the
line positions in the 2012 edition of the HITRAN database \cite{jt557}. The
standard deviation of the relative difference between the experimental and
HITRAN OH line positions is 0.0076 cm$^{-1}$. Since HITRAN does not contain data
on the pressure shift of the OH lines, we estimate the shift to be 10\%\ of the
pressure broadening at atmospheric pressure, i.e. 0.007 \cm. Thus we estimate the frequency
accuracy of the spectrum is  0.01 cm$^{-1}$. The high-temperature
spectrum is averaged 20 times and normalized to a background spectrum measured
when the flame is off. The baseline is additionally corrected for slowly varying
etalons fringes. 

We note that the influence of broadband flame emission can be neglected since
the probablity of emission into the cavity mode is low and the cavity thus acts
as an effective filter. Moreover, the collimator for coupling the cavity
transmitted light into the fiber does not face the flame and is placed few tens
of cm away from the flame, where the intensity of emission is already very low.

\section{Cavity-enhanced absorption spectrum}

\begin{figure}
\centering
\scalebox{0.6}{\includegraphics[scale=0.8]{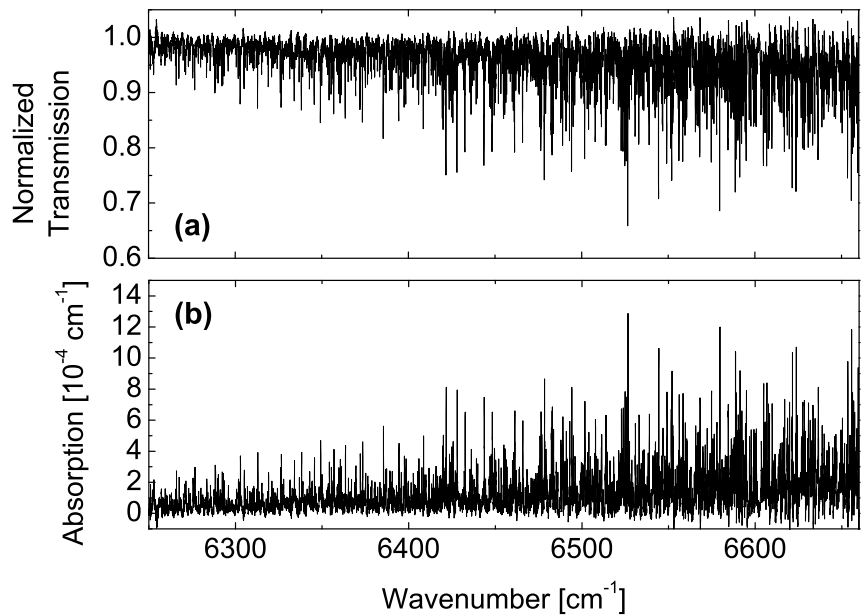}}
\caption{(a) Normalized cavity-enhanced transmission spectrum in
  the flame at atmospheric pressure, [H$_2$O] = 17\%\ and T = 1950 K.
  (b) The corresponding absorption spectrum.}
\label{f:1}
\end{figure}

The normalized transmission spectrum measured in the flame is shown in
Fig.~\ref{f:1}(a). To extract the absorption coefficient from this spectrum,
 we use the model for the transmission, $ I_T$, given by Foltynowicz {\it et al.} \cite{UMU6}
\begin{equation}
 I_T(\nu) = \frac{(1-r)^2 \exp[-\alpha(\nu)L]}{1 +r^2\exp[-2\alpha(\nu)L]
 -2r\exp[-\alpha(\nu)L]\cos[2\phi(\nu)L+\varphi(\nu)]}~,
\end{equation}
where $L$ is the interaction length between the light and the sample
(i.e. the flame diameter), $r$ is the frequency-dependent
intensity reflection coefficient of the cavity
mirrors, determined experimentally by cavity ringdown, $\alpha$ and
$\phi$ are the molecular absorption and dispersion coefficients, respectively,
and $\varphi$ is the
round-trip phase shift in the cavity. The round-trip intracavity phase
shift is equal to a multiple of $2\pi$ for comb lines locked to the
centers of the corresponding cavity modes. Because of the intracavity
dispersion, caused by the cavity mirror coatings as well as the
gas sample inside the cavity, the cavity modes are not equally spaced
and only the comb lines around the locking points are exactly on
resonance with their corresponding cavity modes \cite{UMU6}. However, because of the low
cavity finesse, the relative comb-cavity offset is small in the entire
spectral range of the comb, and the intracavity phase shift can be set to $2\pi$, or zero. To extract the absorption
coefficient from Eq. (1) we also neglect the molecular dispersion,
since then the equation can be solved analytically. This approximation
gives correct values for on-resonance absorption coefficients, since
molecular dispersion is equal to zero at these frequencies.

The absorption spectrum obtained using the analytical solution to Eq.
(1) with molecular and cavity dispersion put to zero is plotted in Fig.~\ref{f:1}(b).
The noise on the baseline is $5 \times
10^{-7}$ cm$^{-1}$, which translates into a signal-to-noise ratio (SNR) of
2400 for the strongest lines. The negative absorption values and the
slight line asymmetry at frequencies above 6550 cm$^{-1}$ are caused by the
neglected comb-cavity offset, which increases away from the locking
points \cite{UMU6}.

\section{Experimental water line list}

The center frequencies of absorption lines are found
by taking the first derivative of the absorption spectrum [Fig.~\ref{f:1}(b)] and finding the points where it crosses zero. Most of the absorption lines are
water transitions but the spectrum contains also several OH
transitions \cite{UMU10}. The OH transition frequencies, identified
using the 2012 edition of the HITRAN database \cite{jt557}, as well as
water lines less than 0.02 \cm\ away from an OH line, are removed from
the list. The precision of the center frequencies is below 0.0005 \cm\ for most lines, limited by the line width of the molecular lines (0.7 \cm) and the SNR of up to 2400. The accuracy is limited to 0.01 \cm\ by the HeNe wavelength calibration. It should be emphasized that positions of overlapping water lines,
i.e. those separated by less than the line width, cannot be identified
using this method. Note also that the transition frequencies are at atmospheric
 pressure.

The experimental line intensities, $S$, are calculated from the value
of absorption $\alpha_{\rm max}$ corresponding to each center frequency, using
\begin{equation}
 S = \frac{\alpha_{\rm max}}{n_T\chi_{\rm max}} ,
\end{equation}
where $\chi_{\rm max}$ is the peak (on-resonance) value of the Voigt profile (in cm),
and $n_T$ is water density at the temperature $T$ (equal to $6.4
\times 10^{17}$ molecule/cm$^3$ for $T = 1950$ K and [H$_2$O]=17\%).
Since no data exists for the pressure broadening parameter of water at
these temperatures, we assume a Lorentzian half width of 0.027 \cm for all lines, as it matches relatively well to the data.
The Doppler half width varies from 0.0237 to 0.0253 \cm across the
spectrum. The experimental line list contains 2417 lines; it is
plotted in Fig.~\ref{f:3}
and given in the supplementary information.
The lowest line intensity that can be identified
is $10^{-25}$ cm/molecule, limited by the SNR in the
spectrum. The uncertainty in the intensity is  6\% for
the strongest lines, limited
mainly by the uncertainty in the water
concentration, and increases for weaker lines
because of the lower SNR.

\begin{figure}
\centering
\scalebox{0.6}{\includegraphics[scale=0.8]{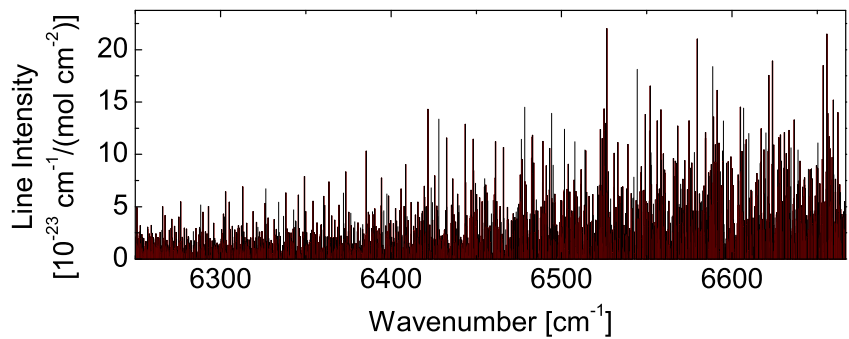}}
\caption{Line intensities of water transitions at 1950 K retrieved from the measurement in the flame.}
\label{f:3}
\end{figure}

%\section{Comparisons of measured and calculated line lists}

\begin{figure}
\centering
\scalebox{0.6}{\includegraphics[scale=0.8]{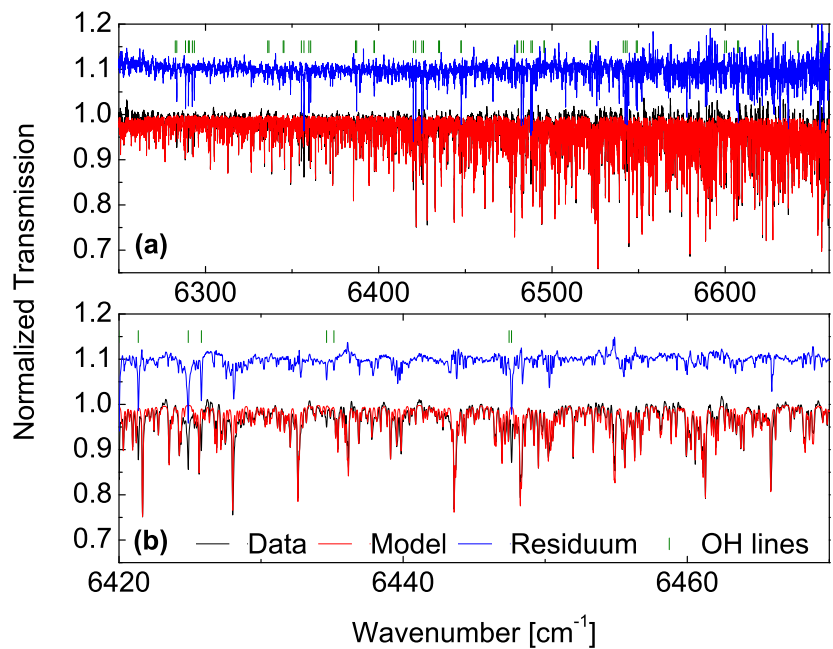}}
\caption{Comparison between the measurement (black) and the spectrum calculated using the experimental line list (red).
Residuum is shown in blue, vertically offset for clarity. The frequencies of the strongest OH transitions are marked by green lines.
(a) Full range. (b) Enlarged section showing the general good agreement between the data and the model.}
\label{f:4}
\end{figure}

To illustrate the accuracy of the experimental line list, Fig.~\ref{f:4}(a)
shows a comparison between the experimental
transmission spectrum (black) and a spectrum simulated using Eq. (1)
with the experimental line list, Voigt  profiles with a
Lorentzian half width of 0.027 \cm and the intracavity round-trip
phase shift put to zero (red). The difference between the measurement and
the model is plotted in blue (vertically offset for clarity). The green vertical lines mark the positions of OH lines. Figure~\ref{f:4}(b)
shows a zoom of the spectrum around one
of the locking points, i.e. where the comb-cavity offset is zero. The structure in the residuum at the frequencies marked in green comes from the OH lines that are removed from the line list. The discrepancies visible at other frequencies are caused mainly by an incorrect  Lorentzian width and by the remaining water lines that could not be taken into account in the line list because of their strong overlap with other lines. The amplitude of the residual increases for higher wavenumbers since the comb-cavity offset increases away from the locking points.

\section{Line assignment}
%\red{I rearanged this section, we should start by describing how the entire assignment has been done, then show the figure illustrating multiple assignments, and finally the figure comparing the calculated line list to our data and to HITRAN. The text needs fixing in multiple places.}
The spectral analysis of the experimental line list was performed using the recently computed POKAZATEL
%\red{explain what it stands for}
 hot line list
\cite{jtpoz} with energies replaced by empirical energy levels
\cite{jt539}. These empirical levels come from the recent
IUPAC-sponsored study of water spectra \cite{jt562} in which the
MARVEL (measured active vibration-rotation energy levels)
\cite{jt412,12FuCs.method} procedure was used to invert measured line
frequencies to obtain empirical energy levels; below, this line list is refered
to as MARVEL-POKAZATEL.
% The line positions of the calculated line list match well to the positions of the measured lines.
Using the  MARVEL-POKAZATEL line list to generate a water absorption
spectrum at 1950 K in the experimental wavelength range leads to about 20~000 lines with
intensities between 10$^{-22}$ and 10$^{-25}$ cm/molecule, which
corresponds to the experimental
intensity range.

%\red{Add a sentence showing we are changing gears?}

The first step in the spectral analysis involved making a so-called
trivial assignment, that is identifying lines for which both upper and lower
state energy levels as well as frequencies are already known empirically. 
%If the upper and
%the lower levels of variational transition are empirically known
%an empirical
%\red{please do not use 'experimental' when writing about MARVEL, because that is confused with our experimental line list}
%frequency can be calculated.  
We assumed a match
between MARVEL and experimental frequencies if they differ by less than
about 0.03 \cm, which reflects the average line width at
atmospheric pressure of about 0.07 \cm and the accuracy of hot
MARVEL levels which is about 0.02 \cm.  About 1900 lines were trivially
assigned by this method with about half of them being identified as blends,
that is two or more MARVEL transitions associated with one experimental line.

The second step in the spectral analysis was assignment of the
remaining stronger lines that were not trivially using the POKAZATEL line list.
The trivially-assigned lines were also considered to be unassigned
during this second step if their calculated intensity was less than
half of the measured one. We only considered stronger theoretical
lines with calculated intensities higher than 10$^{-23}$ cm/molecule.
We concentrated on identifying those transitions which involved upper
levels with quantum numbers close to MARVEL levels. This allowed us to
estimate the expected observed minus calculated residue for the
upper level using the difference between known MARVEL levels and
levels predicted by the original POKAZATEL line list.
%\red{This  sentence is not clear, what is the obs-calc residue?}
This method
of tracking states with nearby quantum numbers is sometimes known as
the method of branches \cite{jt205,jt633}; it works best for the
relatively rare, stronger experimental lines. Use of this method led
to the assignment of about 300 further lines,  providing
information on 126 new energy levels. We note that this method also implicitly provides
the vibrational quantum numbers for the newly assigned levels.
%\red{sy: I was wondering if we could include a table with the new 127 energy level. This is important data, I think it deserves to be included in the main text. We are currently not planning to have it as part of the supplementary material, are we? } 
As a result 2030 lines out of the observed 2417 lines from the experimental line list, i.e. almost 84\%, are assigned. 
When blends are considered, the total number of assigned lines is 2937. 
Some of the unassigned lines may be attributable to other species present in the flame, 
such as CO or CO$_2$. For example the spectrum covers the region of the CO second overtone. However,
attempts to model CO and CO$_2$ lines in this region did not produce clear matches.
%\red{( $\gets$ Is my interpretation of the following sentence correct? \textit{These assignments are associated with 2937 separate transitions.})}

\begin{figure}
\centering
\scalebox{0.66}{\includegraphics[scale=0.8]{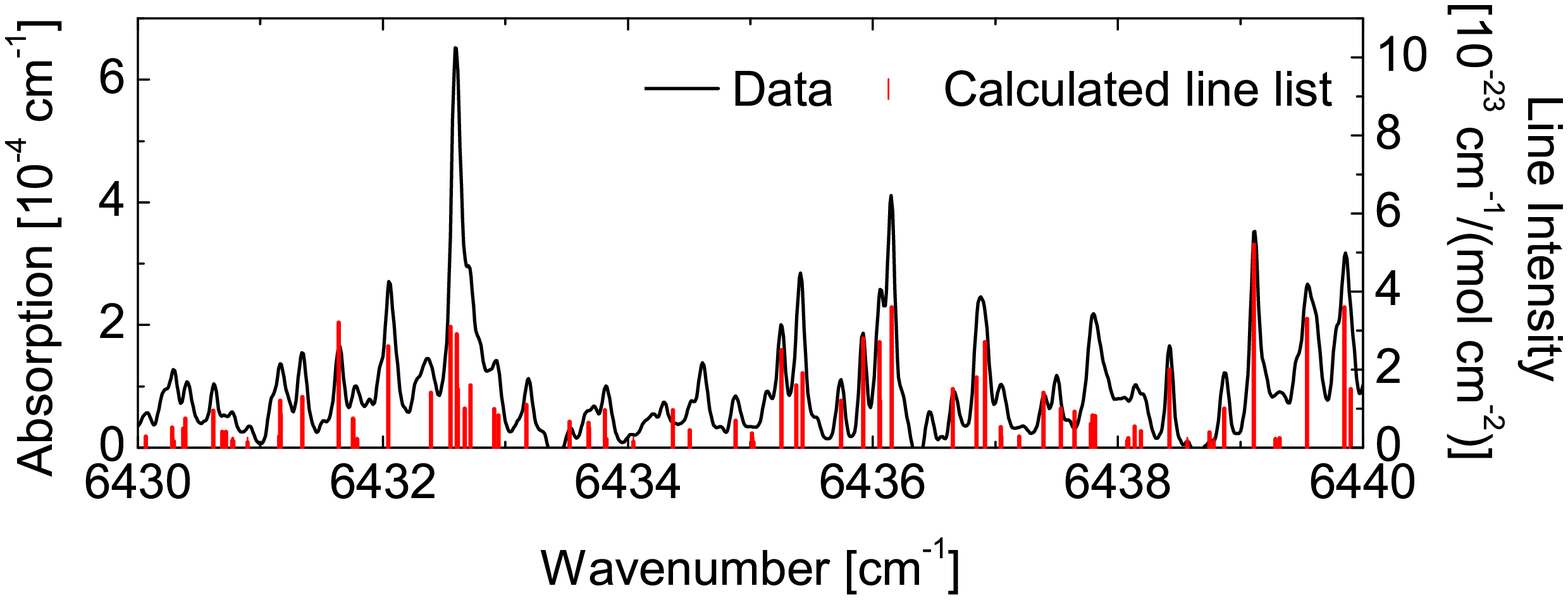}}
\caption{Comparison of the measured spectrum (black, left axis)
and a stick spectrum of the assigned calculated lines (red, right axis) for a small frequency window.}
\label{f:6}
\end{figure}

Figure~\ref{f:6} shows 10 \cm\ of the measured spectrum around 6435 \cm\ compared to the calculated line list and illustrates situations when multiple lines are associated with a single observed feature.
For example, the strong, broad feature at 6432.6 \cm\ consists of at least three actual transitions, however in the
experimental line list these transitions are represented by a single
line at the strongest peak at 6432.595 \cm.  We assigned 3 MARVEL
lines to this feature; these lines have a summed intensity about 65\%\ of
the measured integral intensity of the feature, meaning it could be hiding
further lines.  The line at 6439.858 \cm\
appears to be isolated but in fact has a double
assignment with the difference between the two MARVEL frequencies of
0.05 \cm; these lines model essentially the whole intensity of this
feature. Single MARVEL assignment means that any other nearby
lines are at least about an order-of-magnitude weaker.

%Figure~\ref{f:5} shows a comparison between the experimental data
%(black) and the cavity-enhanced spectra calculated using the
%experimental line list (red, same as in Fig.~\ref{f:3}) and using the
%calculated line list (blue, 1950 K, $J = 0 \rightarrow 25$), both
%assuming a Lorentzian half with of 800 MHz.
%\red{We need to add comparison to HITRAN and a discussion about the discrepancies in line intensities.}

%\begin{figure}
%\centering
%\scalebox{0.6}{\includegraphics[scale=0.8]{Fig5.eps}}
%\caption{Comparison between the data (black) and the spectra calculated using %(a) the experimental line lists (red), (b) the calculated line list (green), %and (c) the line parameters from the HITRAN database (blue).}
%\label{f:5}
%\end{figure}

The calculated line list contains many
more weak lines than the experimental line list. These weak lines
overlap with more intense water lines and therefore cannot be
identified from the experimental spectrum. However, the contribution
of these weak lines does not explain many apparent line strength discrepancies
between the two line lists.
%\red{Any thoughts on what might explain these differences?}
The task of predicting line intensities for water using {\it ab initio}
procedures is under constant review \cite{jt687} and work is
currently in progress at UCL to further improve the water dipole
surface. Progress on this will be reported elsewhere.

A full list of the experimental lines with assignments are given in the
supplementary data. This list specifies whether the line was
trivially assigned using MARVEL or is associated with a new energy
level. We note that the short spectral range and density of lines
meant that these new energy levels are not generally confirmed
by combination differences.

The spectrum contains transitions from 136 bands, of which 45 contain only a
single transition. Table 1 shows a summary of the main bands observed in this
spectrum. Only about 20\%\ of the
observed lines involve transitions from the vibrational ground state
with most corresponding to hot bands. Transitions involve a large
number of rotational states with $J$ up to 27. 

Table 2 presents our newly determined energy levels. Differences between these
and the values predicted by the POKAZATEL line list are also given. The small value of these differences
and their smooth behavior within a given vibrational state lends confidence to our
new assignments. Note that these energy levels are derived from measurements
made at atmosheric pressure and therefore will include small contributions due to
the pressure shift.
For comparison, Zobov {\it et al} \cite{jt437} analyzed a hot emission spectrum
recorded using an oxy-acetylene welding
torch and a Fourier transform spectrometer. In the 6250 -- 6670 \cm\ region 
considered here that spectrum contains a similar number of lines to the
current one, albeit line positions were only determined to about 0.02 \cm.
Zobov  {\it et al} give 840 line assignments in this region which can be
compated with 2937 here. This means that more than 70\%\ of the line assignments
given here are actually new.

\begin{table}
        \caption{Summary of vibrational bands with $N$ lines
assigned to them. Only bands for which $N \geq 10$ are given.}
        \begin{tabular}{crlcr}
                \hline\hline
 band  &    $N$  && band   &   $N$ \\
\hline
021-000 &187 &    &                 032-011 & 40\\        
101-000 &182 &    &                 141-120 & 31\\        
200-000 &142 &    &                 400-200 & 29\\        
120-000 & 75 &    &                 310-110 & 28\\        
002-000 & 36 &    &                 211-011 & 27\\        
011-000 & 27 &    &                 220-001 & 27\\        
040-000 & 10 &    &                 042-021 & 23\\        
111-010 & 93 &    &                 022-100 & 20\\        
301-200 & 68 &    &                 121-001 & 19\\        
102-100 & 65 &    &                 103-002 & 19\\        
211-110 & 65 &    &                 301-101 & 19\\        
102-001 & 64 &    &                 140-020 & 18\\        
400-101 & 56 &    &                 230-011 & 17\\        
310-011 & 55 &    &                 410-111 & 17\\        
210-010 & 49 &    &                 221-120 & 15\\        
061-040 & 46 &    &                 221-200 & 15\\        
131-110 & 43 &    &                 021-010 & 13\\        
022-001 & 42 &    &                 220-100 & 12\\        
130-010 & 41 &    &                 311-210 & 12\\        
121-100 & 40 &    &                 220-020 & 11\\        
                \hline\hline                              
        \end{tabular}                                     
 \end{table}        
                                      
\begin{table}
        \caption{Newly determined energy levels for H$_2$$^{16}$O. Observed -- calculated (o-c) values, in \cm, are relative
to POKAZATEL line list}
{\tiny
        \begin{tabular}{ccrrcccrrcccrr}
                \hline\hline
$J$K$_a$K$_c$&$v_1v_2v_3$& E / \cm  & o-c&&  $J$K$_a$K$_c$&$v_1v_2v_3$& E / \cm  & o-c&&$J$K$_a$K$_c$&$v_1v_2v_3$& E / \cm  & o-c\\
\hline
14  1 13&	061&	15196.170&	-0.044  & &     8  3  5&	061&	13975.084&	-0.021 &&          16  7  9&	201&	14398.683&	-0.037 \\     
15 12  3&	002&	11944.059&	-0.038  & &     9  3  7&	061&	14184.268&	-0.033 &&          17  0 17&	201&	13436.679&	-0.104  \\      
19  9 10&	002&	12947.908&	-0.044  & &     6  2  4&	071&	14768.262&	-0.013 &&          18  4 14&	201&	14817.74&	-0.064  \\        
 8  1  8&	004&	15233.158&	-0.02   & &    15  9  6&	101&	11077.237&	-0.145 &&          18  7 11&	201&	15211.965&	-0.168  \\       
16  2 15&	021&	9813.798&	-0.028   & &   17  8  9&	101&	11645.203&	-0.068 &&          10  9  2&	210&	11277.802&	-0.033  \\       
16  2 15&	022&	13412.179&	-0.057  & &    18  9 10&	101&	12258.106&	-0.048 &&          11 11  0&	210&	11991.199&	 0.063  \\       
 9  4  6&	031&	9852.648&	-0.099   & &   23  5 19&	101&	13720.971&	-0.034 &&          12  9  4&	210&	11831.292&	-0.027  \\       
12  5  7&	031&	10861.193&	 0.169  & &    26  6 20&	101&	15782.146&	-0.074 &&          12 10  3&	210&	12013.462&	 0.018  \\       
19  7 13&	031&	13925.583&	-0.04   & &    15  6  9&	102&	14189.387&	-0.084 &&          12 12  1&	210&	12543.449&	 0.067  \\       
13  7  6&	032&	15168.281&	-0.035  & &    14  8  7&	111&	12198.087&	-0.113 &&          13 10  3&	210&	12320.932&	 0.011  \\       
12 10  3&	040&	10359.146&	-0.046  & &    15  8  7&	111&	12549.411&	-0.138 &&          13 11  2&	210&	12587.499&	 0.035  \\       
11  5  6&	041&	12161.521&	 0.007  & &    16  8  9&	111&	12922.244&	-0.065 &&          13 12  1&	210&	12854.881&	-0.059  \\       
11  6  6&	041&	12401.372&	-0.056  & &    20  6 14&	111&	14382.203&	-0.028 &&          15  5 10&	210&	12071.533&	-0.053  \\       
12  5  7&	041&	12442.132&	-0.024  & &    21  8 14&	111&	15088.616&	-0.031 &&          16  4 13&	210&	12132.162&	-0.06   \\       
12  8  4&	041&	13250.26&	 0.007   & &   13  4  9&	112&	14878.042&	 0.042 &&          17  4 13&	210&	12698.242&	-0.004  \\       
13  7  7&	041&	13263.805&	-0.121  & &    16  9  8&	120&	11396.542&	-0.012 &&          17  5 12&	210&	12867.315&	-0.027  \\       
14  7  7&	041&	13594.493&	-0.036  & &    20  8 13&	120&	12913.391&	 0.015 &&          18  2 17&	210&	12287.241&	-0.051  \\       
16  2 14&	041&	13214.275&	 0.081  & &    17  3 15&	121&	13851.555&	-0.026 &&          18  3 16&	210&	12598.021&	-0.041  \\       
19  3 17&	041&	14350.205&	-0.006  & &    19  2 18&	121&	14244.954&	 0.012 &&          18  4 15&	210&	12871.799&	-0.078  \\    
 5  3  3&	051&	11980.727&	-0.009  & &    20  1 19&	121&	14622.29&	-0.126 &&          19  1 18&	210&	12646.72&	-0.057  \\    
 6  3  3&	051&	12129.054&	 0.0    & &     9  8  1&	130&	10759.505&	-0.006 &&          19  2 17&	210&	12974.57&	-0.05   \\    
 6  4  2&	051&	12366.269&	-0.082  & &    10  8  3&	130&	11001.243&	-0.029 &&          19  3 16&	210&	13264.616&	-0.067  \\    
 7  2  5&	051&	12135.336&	-0.024  & &    10  9  2&	130&	11185.909&	 0.003 &&          20  1 20&	210&	12624.253&	-0.056  \\    
 7  5  3&	051&	12805.967&	-0.004  & &    11  8  3&	130&	11265.526&	-0.018 &&          20  2 19&	210&	13022.745&	-0.067  \\    
 8  1  8&	051&	11996.135&	 0.018  & &    12  8  5&	130&	11551.968&	-0.016 &&          21  0 21&	210&	12996.448&	-0.028  \\    
 8  4  4&	051&	12730.885&	-0.023  & &    12  9  4&	130&	11736.443&	-0.006 &&          21  1 20&	210&	13415.475&	-0.061  \\    
 9  0  9&	051&	12164.81&	-0.022   & &   13  8  5&	130&	11860.261&	-0.03  &&          16  9  7&	211&	16439.527&	-0.053  \\    
 9  5  5&	051&	13215.983&	-0.035  & &    13  9  4&	130&	12044.351&	-0.024 &&          13  4  9&	220&	12824.952&	-0.043  \\    
11  2 10&	051&	12914.709&	-0.02   & &    14  6  9&	130&	11643.342&	-0.05  &&          10  9  2&	300&	12856.023&	 0.029  \\    
11  7  5&	051&	14302.282&	 0.044  & &    14  8  7&	130&	12189.994&	-0.019 &&          13  8  5&	300&	13510.443&	-0.045  \\    
12  1 11&	051&	13157.177&	 0.015  & &    12  2 11&	140&	11582.593&	-0.011 &&          15  3 12&	300&	13515.85&	-0.053  \\    
12  7  5&	051&	14586.049&	-0.016  & &    11  3  9&	141&	15061.568&	-0.033 &&          16  2 15&	300&	13390.283&	 0.132  \\    
13  3 11&	051&	13756.382&	-0.046  & &    12 11  1&	200&	10531.411&	-0.094 &&          16  4 13&	300&	13856.623&	-0.069  \\    
 2  0  2&	061&	12656.003&	-0.004  & &    15  9  6&	200&	11086.936&	-0.033 &&          16  4 13&	300&	13856.632&	-0.059  \\    
 2  2  0&	061&	12923.88&	 0.007   & &   16  6 10&	200&	10935.016&	-0.037 &&          16  5 12&	300&	14057.023&	-0.068  \\    
 3  1  3&	061&	12787.646&	 0.035  & &    16 10  6&	200&	11651.177&	-0.035 &&          18  3 15&	300&	14581.009&	-0.093  \\    
 3  3  1&	061&	13245.215&	 0.01   & &    16 13  4&	200&	12398.076&	 0.03  &&          18  7 12&	300&	15158.842&	-0.051  \\    
 4  0  4&	061&	12813.251&	-0.015  & &    18  5 13&	200&	11650.563&	-0.006 &&          19  5 14&	300&	15399.793&	-0.111  \\    
 4  1  3&	061&	12933.841&	-0.036  & &    18  9 10&	200&	12267.551&	 0.049 &&          12 10  3&	310&	15298.892&	-0.026  \\    
 4  2  2&	061&	13095.518&	 0.001  & &    13  4  9&	201&	13052.721&	-0.041 &&          14  3 12&	310&	14588.529&	-0.015  \\    
 4  3  2&	061&	13342.05&	-0.009   & &   15  7   9&	201&	14005.593&	-0.075 &&          17  3 14&	310&	15807.947&	-0.034  \\
                \hline\hline                              
        \end{tabular}          }                           
 \end{table}        

\section{Conclusion}

A near-infrared absorption spectrum of water recorded in a flame at 1950 K using 
cavity-enhanced optical frequency comb-based Fourier transform spectrometer is shown to be a
rich source of information on water transitions. About 85\%\ of the
lines observed in the spectral region 6250 -- 6670 \cm are assigned using both previous information on empirical
energy levels and by comparison with a new, variational line list.
Many of the experimental lines are assigned to multiple transitions.
The majority of the assigned lines are actually associated with hot
bands.  These new data will form
part of the input for the update of MARVEL energy levels for H$_2$$^{16}$O,
which is currently in progress \cite{jtwaterupdate}.
% \red{Some of the unassigned lines could be also attributed to other species contaminating our experiment, such as CO or CO$_2$}.

\section*{Acknowledgements}

This work was supported by NERC under various grants. A.F. acknowledges support from the Swedish Research Council (2016-03593), the Knut and Alice Wallenberg Foundation (2015.0159), and the Swedish Foundation for Strategic Research (ICA12-0031). F.M.S. acknowledges financial support by the Swedish Energy Agency (36160-1), the Kempe Foundations (JCK-1316) and the Swedish strategic research program Bio4Energy. A.A.K and N.F.Z. acknowledge support by State Project IAP RAS No. 0035-2014-009.

\newpage
\bibliographystyle{elsarticle-num}
%
%
%\bibliography{journals_phys,jtj,UmU,methods,H2O,h216o,diatomic,waternewsources,CO2,hodges,linelists,atmos}

\end{document}